\documentclass[runningheads]{llncs}

\usepackage[T1]{fontenc}
\usepackage{graphicx,verbatim}
\usepackage{tabularx}
\usepackage{siunitx}
\usepackage{marvosym}
\usepackage{amsmath,amssymb}
\usepackage{makecell}
\usepackage{multirow}
\usepackage{booktabs}
\usepackage{cite}
\usepackage[colorlinks=true,citecolor=blue,linkcolor=blue,urlcolor=blue]{hyperref}
\usepackage{makecell}
\usepackage{xspace}
\newcommand{\methodname}{SEER\xspace}

\usepackage{array}
\newcolumntype{I}{!{\vrule width 3pt}}
\newlength\savedwidth

\newlength\savewidth
\newcommand\shline{\noalign{\global\savewidth\arrayrulewidth
\global\arrayrulewidth 1.25pt}%
\hline
\noalign{\global\arrayrulewidth\savewidth}}

\begin{document}

\title{Skill-Evolving Grounded Reasoning for Free-Text Promptable 3D Medical Image Segmentation}
\titlerunning{\methodname for Free-Text Promptable 3D Medical Image Segmentation}

\author{
    Tongrui Zhang\inst{1} \and 
    Chenhui Wang\inst{1} \and 
    Yongming Li\inst{1} \and 
    Zhihao Chen\inst{1} \and 
    Xufeng Zhan\inst{2} \and 
    Hongming Shan\inst{1} 
}


\authorrunning{T. Zhang et al.}

\institute{
    Fudan University \and University of Science and Technology of China \\
}
  
\maketitle             

\begin{abstract}
Free-text promptable 3D medical image segmentation offers an intuitive and clinically flexible interaction paradigm. However, current methods are highly sensitive to linguistic variability: minor changes in phrasing can cause substantial performance degradation despite identical clinical intent. 
Existing approaches attempt to improve robustness through stronger vision-language fusion or larger vocabularies, yet they lack mechanisms to consistently align ambiguous free-form expressions with anatomically grounded representations. We propose \textbf{S}kill-\textbf{E}volving ground\textbf{E}d \textbf{R}easoning (\textbf{\methodname}), a novel framework for free-text promptable 3D medical image segmentation that explicitly bridges linguistic variability and anatomical precision through a reasoning-driven design. 
First, we curate the \methodname-Trace dataset, which pairs raw clinical prompts with image-grounded, skill-tagged reasoning traces, establishing a reproducible benchmark.
Second, \methodname constructs an evidence-aligned target representation via a vision-language reasoning chain that verifies clinical intent against image-derived anatomical evidence, thereby enforcing semantic consistency before voxel-level decoding. 
Third, we introduce \methodname-Loop, a dynamic skill-evolving strategy that distills high-reward reasoning trajectories into reusable skill artifacts and progressively integrates them into subsequent inference, enabling structured self-refinement and improved robustness to diverse linguistic expressions.
Extensive experiments demonstrate superior performance of \methodname over state-of-the-art baselines. Under linguistic perturbations,  \methodname reduces performance variance by 81.94\% and improves worst-case Dice by 18.60\%. Project resources are available at: \url{https://seer-medseg.github.io}.

\end{abstract}

\section{Introduction}
Text-guided 3D medical image segmentation~\cite{zhao2025large,zhao2025foundation,zhao2025boltzmann,xin2025text3dsam} allows users to specify targets via predefined class names, offering a more intuitive interface for clinical workflows compared to traditional point- or bounding-box-based models~\cite{kirillov2023segment, ma2024segment, ma2025medsam2, liu2026medsamagent, chen2023ascon}. Yet, the dominant language signal available in real-world care is not a standardized label set but free-text: unstructured, free-form language originating from imaging orders, clinical questions, or radiology reports. Free-text promptable segmentation broadens clinical applicability for diverse diagnostic scenarios and also serves as a natural interface for future AI agent integration.

However, directly conditioning segmentation models on free-text introduces severe challenges regarding linguistic variability, as clinical text frequently contains abbreviations, synonymous phrasing, and institution-specific conventions~\cite{larson2025assessing}. While recent free-text promptable segmentation work attempts to alleviate this through larger training corpora, stronger vision-language fusion~\cite{rokuss2025voxtell_cvpr2026}, or more complex multi-agent systems~\cite{liu2025medsam3delvingsegmentmedical}, they largely overlook that standard natural language embedding spaces are not inherently aligned with complex clinical terminology. Clinically equivalent synonyms or shorthand frequently map to disparate embedding neighborhoods~\cite{deng2026knowledge}, resulting in drastically different conditioning signals and ultimately unstable segmentation masks. Overcoming this instability requires moving beyond direct text-to-feature mapping; it necessitates an intermediate step of explicit reasoning to disambiguate the raw text and resolve the underlying clinical intent before executing the task.

Yet, pure language reasoning is insufficient for medical imaging; it must be heavily grounded in the patient's visual evidence~\cite{wu2025towards,yaomedical,lin2026models}. A dependable text-guided segmentation system needs to comprehend the unstructured clinical request and align it with the anatomical evidence~\cite{lee2025unified,zhu2026medeyes,zhu2026medsynapsevbridgingvisualperception}. Rather than relying on rigid text-to-text synonym normalization, this alignment is best achieved by decomposing the vision-language reasoning into granular, modular steps, known as skills. By applying skills (e.g., synonym normalization or spatial relation reasoning) that cross-reference the textual prompt with visual features or volume renderings, the system can synthesize an explicit, evidence-aligned target representation. This provides the downstream segmentation backbone with a canonical signal, substantially mitigating instability from diverse linguistic expressions.

To this end, we propose a \textbf{s}kill-\textbf{e}volving ground\textbf{e}d \textbf{r}easoning (\methodname) framework, 
which bridges variable free-text prompts to robust 3D medical segmentation by formulating the required vision-language reasoning as explicit, executable skills. When presented with a clinical request and a 3D scan, \methodname leverages these skills to generate a task specification for the segmentation backbone. Furthermore, instead of relying on a static reasoning module, we introduce a dynamic skill-evolving strategy, which continuously distills high-reward grounded reasoning episodes into auditable, reusable skill artifacts. By tracking skill usage and performance gains, \methodname self-refines its reasoning capabilities, progressively enhancing its robustness against complex or unseen linguistic perturbations.

Our contributions are  summarized as follows.
\textbf{(i)} We introduce \methodname, a novel framework that bridges variable free-text clinical prompts to 3D medical segmentation.
\textbf{(ii)} We curate the \methodname-Trace dataset, containing raw clinical prompts with image-grounded and skill-tagged reasoning traces,  to support reproducible research on free-text promptable 3D medical segmentation.
\textbf{(iii)} We formalize grounded visual--language reasoning as explicit skills, enabling the framework to map ambiguous free-text into explicit target representations.
\textbf{(iv)} We propose a dynamic skill-evolving strategy that distills high-reward reasoning episodes into reusable skill artifacts, allowing the model to self-refine and adapt to complex linguistic variations over time.
\textbf{(v)} Extensive experimental results demonstrate that \methodname achieves superior segmentation performance over state-of-the-art baselines and strong transferability across segmentation backbones.

\begin{figure}[t]
\includegraphics[width=\textwidth]{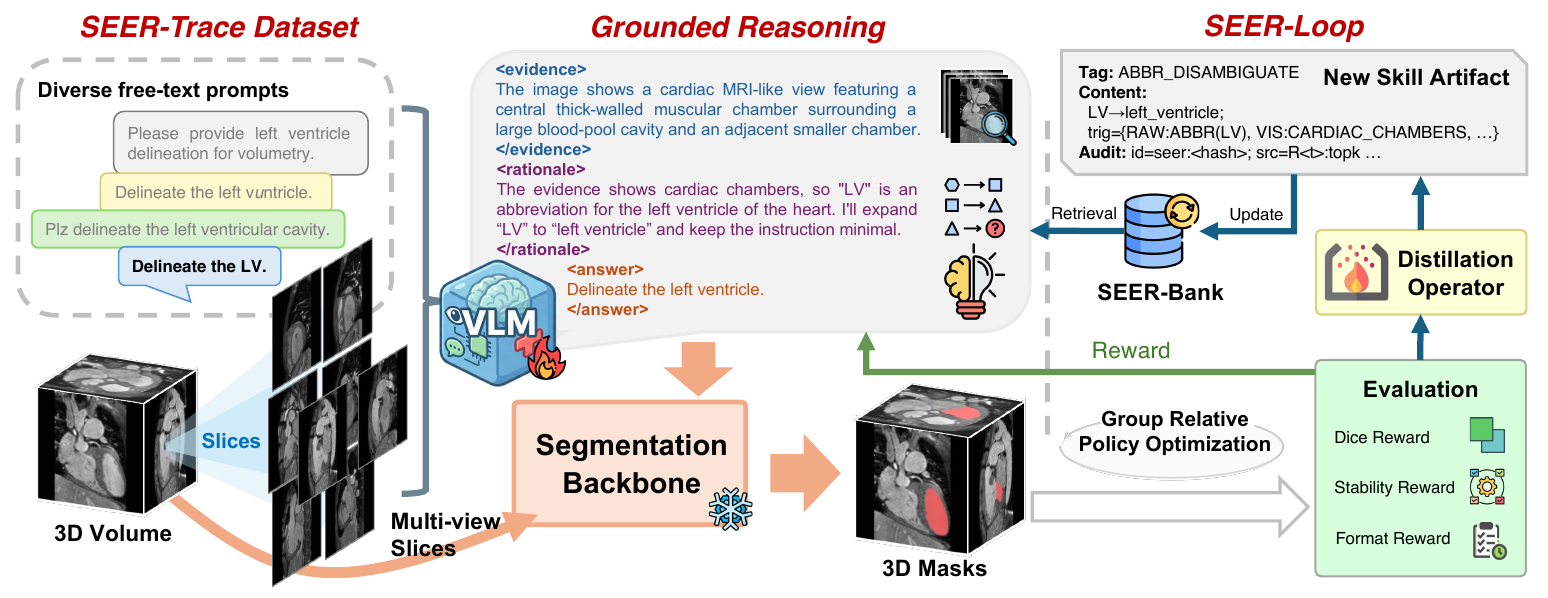}
\caption{Illustration of the proposed \methodname framework. \methodname makes free-text promptable 3D medical segmentation robust by grounding clinical free-text prompts in patient anatomy and evolving reusable reasoning skills.} \label{fig:framework}
\end{figure}

\section{Methodology}
Fig.~\ref{fig:framework} presents an overview of the proposed \methodname framework for free-text
promptable 3D medical segmentation. 
Building upon the supervisory foundation of \textit{\methodname-Trace}, our framework employs \textit{Grounded Reasoning} to accurately align text with multi-view visual evidence; to overcome the limitations of static reasoning, we further introduce a dynamic skill-evolving strategy (\textit{\methodname-Loop}) to distill successful episodes into reusable skills  stored in \methodname-Bank,  thereby establishing a synergistic system that continuously enhances robustness.

\subsection{\methodname-Trace Dataset Curation}

\methodname-Trace is built from established 3D medical segmentation benchmarks~\cite{baid2021rsna,de20242024,hernandez2022isles,ma2024unleashing,pace2024hvsmr} and augments each case with clinician-like prompts and grounded, skill-tagged traces. It essentially aligns three core elements: (i) linguistically diverse prompts, (ii) evidence-aligned intermediate reasoning, and (iii) an executable task specification defined by the downstream segmentation system.

Formally, \methodname-Trace consists of tuples:
\begin{align}
\mathcal{D}_{\textsc{Trace}}
=\big\{(V_i,\; q_i,\; y_i,\; \tau_i)\big\}_{i=1}^{N},
\qquad
y_i=(e_i,r_i,a_i),
\end{align}
where $V_i$ is a 3D volume, $q_i$ is a raw free-text request, and $y_i$ is a structured reasoning response containing \texttt{<evidence>} $e_i$, \texttt{<rationale>} $r_i$, and an executable \texttt{<answer>} $a_i$. Specifically, each raw request $q_i$ is generated under one of eight predefined clinical styles (e.g., radiology note and consult question), producing diverse styles and levels of underspecification. We structure the response $y_i$ as this triplet to explicitly decouple the cognitive steps of intent resolution. Within $y_i$, the evidence $e_i$ enforces perception by isolating observable visual cues from $V_i$, the rationale $r_i$ enforces logical reasoning by linking the ambiguous text $q_i$ to these visual cues, and the answer $a_i$ formulates the final target representation. To augment $y_i$ with modular steps, we also design the skill bank \(\tau_i\), which includes explicit skills such as spatial relation reasoning and synonym normalization.

We render each volume into a multi-view 2D representation $\mathcal{I}_i=\mathcal{R}(V_i)=\{I_{i,m}\}_{m=1}^{M}$. We then use a large language model with constrained prompting to generate $q_i$ and a high-capacity vision-language model (VLM) conditioned on $(\mathcal{I}_i, q_i)$ to synthesize $y_i$ and $\tau_i$, generating intent-preserving explanations and skill tags that are strictly consistent with the volumes and prompts.

To avoid hallucinations, we constrained each request $q_i$ by target aliases, visual priors, and forbidden terms to preserve clinical intent and reduce trivial label leakage. We also checked each candidate $y_i$ for semantic equivalence with the intended target and compatibility with the frozen backbone. Furthermore, we randomly sampled 5\% of tuples for expert audit to verify clinical plausibility. In the expert audit, only 7 out of 1,120 randomly sampled instances were flagged as abnormal, yielding a 0.63\% abnormal rate, which supports the reliability of the full \methodname-Trace dataset. The final \methodname-Trace dataset contains 22,330 multimodal instruction instances from 1,811 cases.

\subsection{Grounded Reasoning}
Building upon the supervised format of \methodname-Trace, this component executes the core reasoning process at inference. 
Rather than solely leveraging a fixed mapping from text input to segmentation type, \methodname actively decomposes complex input prompts into granular, executable cognitive operations, using explicit \emph{skills} such as spatial relation reasoning and synonym normalization.

When processing a clinical request $q$ and rendered views $\mathcal{I}=\mathcal{R}(V)$, the VLM $\pi_{\theta}$ follows the learned structured reasoning format and generates $y=(e,r,a)$:
\begin{align}
\pi_{\theta}(y \mid \mathcal{I}, q, \mathcal{B})
= \pi_{\theta}(e,r,a \mid \mathcal{I}, q, \mathcal{B}),
\end{align}
where $\mathcal{B}$ denotes optional \emph{skill artifacts} retrieved from \methodname-Bank, supplying reusable, high-reward reasoning patterns. To achieve visually grounded reasoning,  three components operate synergistically. First, the model \emph{extracts} anatomical evidence $e$ as a visual grounding constraint in $\mathcal{I}$. Second, it \emph{formulates} a rationale $r$ that logically links the ambiguous free-text request $q$ to this explicit visual evidence, thereby disambiguating the intended target. Finally, based on the visual evidence and rationale, it \emph{synthesizes} an executable answer $a$, which is executed by the frozen segmentation system $\mathcal{S}$, yielding a predicted mask $\widehat{G}$. This structural separation prevents the model from relying on unconstrained latent text conditioning, forcing it to explicitly validate clinical intent against anatomical evidence before generating the final segmentation.

To make our framework robust against diverse linguistic expressions of the same clinical intent, we explicitly optimize the model to produce stable segmentation masks. Let $\Omega(q)$ denote a set of clinically equivalent rephrasings of $q$, \methodname aims to improve both expected segmentation quality and stability under $\Omega(q)$ by optimizing the following objective:
\begin{align}
\mathcal{J}(\theta)=
\mathbb{E}_{(V,q,G)}
\big[&
\mathbb{E}_{q'\sim \Omega(q)}\big[D_{\theta}(V,q',G)\big]
-\lambda 
\mathrm{Var}_{q'\sim \Omega(q)}\big[D_{\theta}(V,q',G)\big]
\big],
\label{eq:robust_obj}
\end{align}
where $D_{\theta}(V,q',G) =\mathrm{Dice} \left(\mathcal{S} \left(V, a_{\theta}(V,q')\right), G\right)$, $G$ is the ground truth segmentation mask, $a_{\theta}(V,q')$ is the generated \texttt{<answer>} for $(V,q')$, and $\lambda$ controls the quality--stability trade-off. 

To effectively maximize $\mathcal{J}(\theta)$, the learning process proceeds in two stages. 
First, supervised fine-tuning (SFT) aligns the VLM with the structural operations of \methodname-Trace. Second, group relative policy optimization (GRPO) further refines the policy via a composite reward function $R$. 
By jointly rewarding the execution Dice score and stability across equivalent queries $q' \in \Omega(q)$, while using a $\beta$-weighted KL penalty to the SFT policy to preserve format compliance, GRPO enforces robust, high-fidelity generation.

\subsection{Dynamic Skill-Evolving Strategy (\methodname-Loop) via \methodname-Bank}
While grounded reasoning resolves ambiguity using existing skills, the linguistic variability in real-world clinical settings is  unbounded. Therefore, we propose a dynamic skill-evolving strategy (\methodname-Loop) implemented via \methodname-Bank. This memory mechanism captures recurring ambiguity patterns as explicit skill artifacts, retrieves relevant skills during generation, and updates the bank over time through experience distillation and downstream feedback; see Fig.~\ref{fig:framework} for the interaction--update loop.

At evolution round $t$, \methodname-Bank stores a set of skill artifacts, $\mathcal{B}^{(t)}=\{s_j=(\kappa_j,  z_j,  \eta_j)\}_{j=1}^{K_t}$, where $\kappa_j$ is a skill tag, $z_j$ is the skill content, and $\eta_j$ records audit metadata. Given rendered views $\mathcal{I}$ and a request $q$, a retrieval operator selects a relevant subset $\mathcal{B}^{(t)}(\mathcal{I},q)\subset \mathcal{B}^{(t)}$ and uses it to guide the model's interaction with the environment by conditioning the generation on retrieved skills:
\begin{align}
\pi_{\theta}(y \mid \mathcal{I}, q, \mathcal{B}^{(t)})
\approx
\pi_{\theta} \big(y \mid \mathcal{I}, q, \mathcal{B}^{(t)}(\mathcal{I},q)\big).
\label{eq:skill_conditioned}
\end{align}
Unlike generic text retrieval, retrieved skills are reusable reasoning strategies. Retrieval ranks skills using both request--metadata compatibility, such as target, tag, and alias overlap, and consistency between stored visual cues and the current multi-view evidence. The selected skills are rendered as structured \texttt{<skill\_bank>} blocks and appended to the VLM input.

Skill evolution is driven by high-reward interaction episodes with the frozen environment. Each episode records the rendered views $I$, request $q$, generated structured response $y=(e,r,a)$, execution result $\widehat{G}=\mathcal{S}(V,a)$, and reward $R$. Let $\Xi^{+}$ denote the set of top-reward episodes. The distillation extracts successful, novel reasoning patterns from $\Xi^{+}$ into new skill artifacts.
To prevent memory bloat and redundancy, we apply two utility-based operations. First, $\mathrm{Dedup}(\cdot)$ groups skills by source, skill tag, and target, merges semantically identical or highly overlapping skills, and keeps the one with higher empirical utility. Second, $\mathrm{Prune}(\cdot)$ further caps the bank size by removing skills with negative or low empirical utility. The bank update is written as:
\begin{align}
\Delta \mathcal{B}^{(t)}=\mathrm{Distill}(\Xi^{+}),
\qquad
\mathcal{B}^{(t+1)}=
\mathrm{Prune}\!\left(
\mathrm{Dedup}\!\left(\mathcal{B}^{(t)}\cup \Delta \mathcal{B}^{(t)}\right)
\right).
\label{eq:skill_update}
\end{align}
The empirical utility used by both $\mathrm{Dedup}(\cdot)$ and $\mathrm{Prune}(\cdot)$ is estimated from each skill's marginal reward gain, which is defined as:
\begin{equation}
\Delta(s)=
\mathbb{E}\big[R \,\big|\, s \in \mathcal{B}(\mathcal{I},q)\big]
-
\mathbb{E}\big[R \,\big|\, s \notin \mathcal{B}(\mathcal{I},q)\big].
\label{eq:skill_gain}
\end{equation}
By tracking these metrics, \methodname-Bank identifies and retains only empirically verified reasoning skills. In effect, this continuous distillation, deduplication, and pruning process stabilizes generation, allowing the framework to self-refine and adapt to complex, unseen linguistic variations over time.

\begin{table*}[t]
\caption{Performance comparison under label and free-text prompting modes.}
\centering
\begin{tabular*}{\textwidth}{@{\extracolsep{\fill}} l l r c r r r}
\shline
\multirow{2}{*}{\textbf{Dataset}}& \multirow{2}{*}{\textbf{Method}} & \textbf{Label Prompting} && \multicolumn{3}{c}{\textbf{Free-text Prompting}} \\
\cline{3-3} \cline{5-7}
 &  & \textbf{Dice} $\uparrow$ && \textbf{Dice} $\uparrow$ & \textbf{Worst Dice} $\uparrow$ & \textbf{Std.} $\downarrow$ \\
\hline
\multirow{6}{*}{\makecell[l]{\textbf{BrainMet}- \\ \textbf{Share}}} 
& SAT~\cite{zhao2025large} & 22.16 && 0.69 & 0.00 & 2.53 \\
& BiomedParseV2~\cite{zhao2025boltzmann} & 18.66 && 2.53 & 0.00 & 7.27 \\
& Text3DSAM~\cite{xin2025text3dsam} & 0.10 && 0.41 & 0.00 & 0.93 \\
& MedSAM3~\cite{liu2025medsam3delvingsegmentmedical} & 11.33 && 16.62 & 10.56 & 5.17 \\
& VoxTell~\cite{rokuss2025voxtell_cvpr2026} & 48.19 && 52.15 & 46.71 & 3.35 \\
& \methodname (\textbf{Ours}) & \textbf{51.70} && \textbf{53.83} & \textbf{51.44} & 1.67 \\
\hline
\multirow{6}{*}{\textbf{PENGWIN}}
& SAT~\cite{zhao2025large} & 96.05 && 0.01 & 0.00 & 0.13 \\
& BiomedParseV2~\cite{zhao2025boltzmann} & 1.35 && 8.53 & 0.00 & 7.50 \\
& Text3DSAM~\cite{xin2025text3dsam} & 24.75 && 0.01 & 0.00 & 0.16 \\
& MedSAM3~\cite{liu2025medsam3delvingsegmentmedical} & 18.26 && 5.75 & 3.67 & 6.40 \\
& VoxTell~\cite{rokuss2025voxtell_cvpr2026} & \textbf{97.59} && 92.26 & 79.34 & 7.49 \\
& \methodname (\textbf{Ours}) & 97.56 && \textbf{97.39} & \textbf{95.47} & 0.98 \\
\shline
\end{tabular*}
\label{tab:sota}
\end{table*}

\section{Experiments}
\subsection{Experimental Setup}

\noindent\textbf{Datasets and prompting protocols.}\quad
We evaluate \methodname on two datasets to comprehensively assess generalization: BrainMetShare~\cite{grovik2020deep} (representing a domain shift involving seen anatomy but unseen institutional sources) and PENGWIN~\cite{liu2025automatic} (serving as a strictly out-of-distribution (OOD) benchmark since its pelvic bone targets are absent from the \methodname-Trace reasoning supervision and target coverage). Each volume is processed into the multi-view 2D representation introduced in \methodname-Trace, paired with diverse, clinician-like free-text prompts. To ensure a comprehensive and fair comparison, we evaluate all baselines under two settings: (i) their natively supported, predefined label prompting interfaces (label prompting mode), and (ii) the realistic, linguistically diverse free-text clinical prompts introduced in \methodname-Trace (free-text prompting mode).

\noindent
\textbf{Implementation details.}\quad 
All experiments are implemented in PyTorch~\cite{paszke2019pytorch} and trained on NVIDIA A100 GPUs. We use DeepSeek-V3.2~\cite{liu2025deepseek} as the high-capacity VLM during \methodname-Trace construction and initialize our vision-language reasoning module with Qwen3-VL-4B-Instruct~\cite{Qwen3-VL}.
During both training stages, the visual encoder is kept frozen while the remaining parameters are fully updated. \methodname-Loop via \methodname-Bank is executed in discrete evolutionary rounds. Following each training phase, we distill high-reward reasoning episodes into reusable skill artifacts, deduplicate the memory bank, and inject these updated skill traces back into the subsequent training stage.

\noindent
\textbf{Evaluation metrics.}\quad 
We report Dice for 3D medical segmentation accuracy. Robustness analyses  under sets of clinically equivalent rephrasings are also reported in terms of standard deviation (Std.) and worst-case Dice (worst Dice).

\begin{figure}[t]
\includegraphics[width=\textwidth]{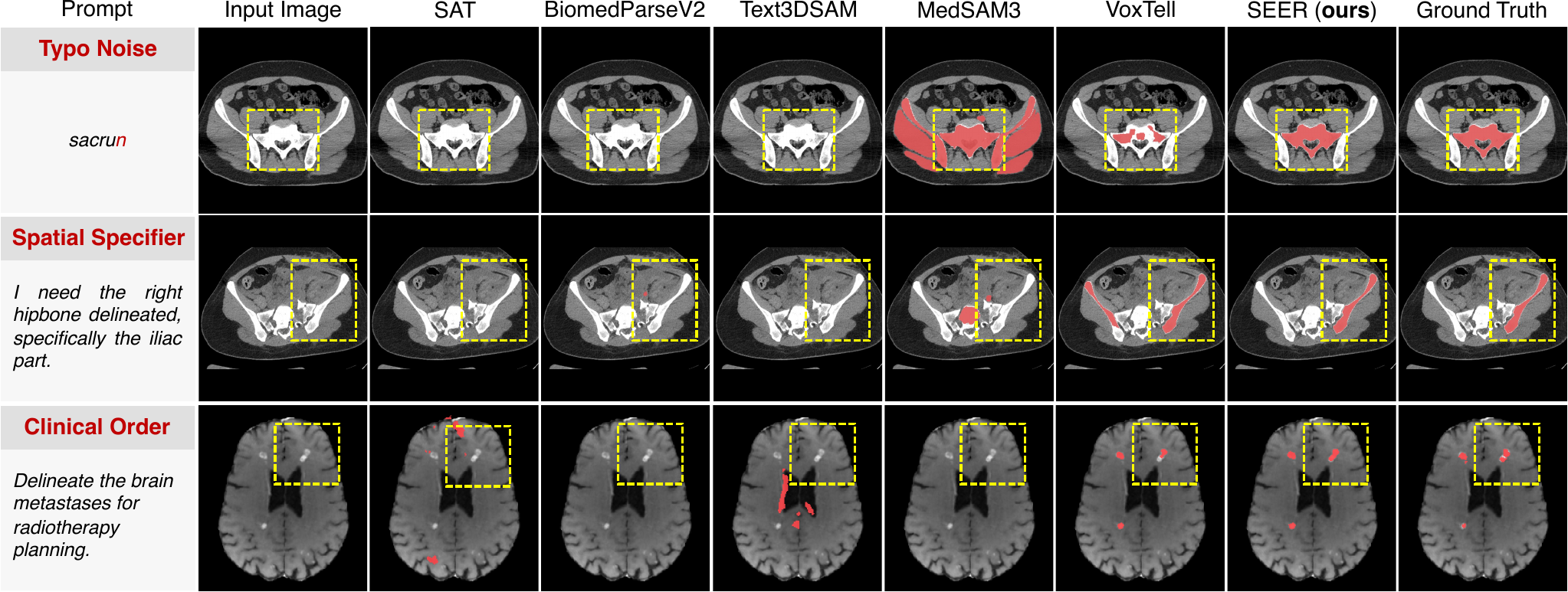}
\caption{Qualitative comparisons under three clinically plausible free-text prompts.} \label{fig:qual_sota}
\end{figure}

\subsection{Comparison Results, Ablation Study, and Generalization}

\noindent\textbf{Quantitative comparison.}\quad
Table~\ref{tab:sota} reports Dice in both label prompting and free-text prompting modes. In label prompting mode, \methodname matches the strongest baseline on PENGWIN and improves on BrainMetShare. In free-text prompting mode, several prior methods collapse to near-zero Dice, reflecting poor robustness to realistic clinical phrasing. Among methods that remain functional, \methodname consistently improves mean Dice while substantially reducing variance and improving worst-case Dice. The consistent gains across both partially and strictly OOD datasets demonstrate that \methodname acquires a new capability—generalizing beyond seen anatomical targets by using grounded reasoning and skill-evolving strategy to resolve ambiguities—rather than merely normalizing long free-text into seen label tokens. 

\noindent\textbf{Qualitative comparison.}\quad
Fig.~\ref{fig:qual_sota} presents representative cases under three clinically plausible free-text prompt variants: \textit{Typo Noise} (spelling perturbation with preserved intent), \textit{Spatial Specifier} (laterality and subregion constraints), and \textit{Clinical Order} (high-level clinical phrasing related to treatment plans). \methodname produces consistent, on-target masks across all variants, reflecting evidence-aligned disambiguation from the image. In contrast, baseline outputs are highly prompt-sensitive in free-text prompting mode, frequently drifting to off-target regions or collapsing to degenerate predictions (e.g., empty/fragmented masks) under minor wording changes.

\noindent\textbf{Ablation study.}\quad 
To evaluate the efficacy of our core architectural contributions, we conducted an ablation study on the strictly OOD PENGWIN dataset. We progressively integrated our proposed modules into the strongest baseline~\cite{rokuss2025voxtell_cvpr2026}. Table~\ref{tab:ablation} shows that  naively prepending a vanilla VLM (Qwen3-VL-4B-Instruct) to parse the text actually degrades performance. This validates our hypothesis that unconstrained, pure-text reasoning without strict anatomical grounding exacerbates instability.
The fine-tuned VLM with grounded reasoning successfully rectifies this, lifting mean Dice to 95.92 and halving the standard deviation compared to the baseline. Finally, the integration of the \methodname-Loop delivers a substantial improvement in robustness. By continuously retrieving empirically verified reasoning artifacts from the \methodname-Bank, this full configuration achieves the highest mean accuracy, raises the worst-case Dice to 95.47, and crushes the standard deviation to 0.98. This near-zero variance indicates that \methodname effectively immunizes the segmentation backbone against linguistic perturbations. 

\begin{table}[t]
\centering
\caption{Ablation study on the strictly OOD PENGWIN dataset.}
\label{tab:ablation}
\begin{tabular*}{\textwidth}{@{\extracolsep{\fill}}lrrr}
\shline
\textbf{Configuration} & \textbf{Dice} $\uparrow$ & \textbf{Worst Dice} $\uparrow$ & \textbf{Std.} $\downarrow$ \\
\hline
Baseline & $92.26$ & $79.34$ & $7.49$ \\
\quad + Vanilla VLM & $84.84$ & $61.90$ & $14.15$ \\
\quad + Fine-tuned VLM w/ Grounded Reasoning & $95.92$ & $88.27$ & $3.84$ \\
\quad\quad \textbf{+ \methodname-Loop (our \methodname)} & $\mathbf{97.39}$ & $\mathbf{95.47}$ & $\mathbf{0.98}$ \\
\shline
\end{tabular*}
\end{table}

\noindent\textbf{Cross-backbone generalization.}\quad 
To evaluate the generalizability of \methodname across different foundational architectures, we replace the downstream segmentation backbone with MedSAM3 while keeping our proposed front-end frozen. 
Specifically, on the strictly OOD PENGWIN dataset, integrating \methodname substantially improves upon the standalone MedSAM3 baseline, increasing the mean Dice from 5.75 (std: 6.40) to 19.97 (std: 13.30) and elevating the worst-case minimum score from 3.67 to 3.94. This suggests that the reasoning capabilities of \methodname are beneficial to other segmentation backbones, consistently elevating zero-shot generalization and worst-case robustness.

\section{Conclusion}
We presented \methodname, a skill-evolving grounded reasoning framework that stabilizes free-text promptable 3D medical segmentation by converting linguistically diverse clinical prompts into a single evidence-aligned, executable task specification for a frozen segmentation system. Across partially and strictly OOD benchmarks, \methodname preserves strong label prompting performance while substantially improving free-text prompting robustness, reducing dispersion under linguistic variability and improving worst-case behavior; ablations further show that grounded visual reasoning and \methodname-Loop contribute complementary gains. Future work should explore volumetric VLM encoders, expand the skill taxonomy, evaluate broader clinical tasks, and carefully audit failure modes to ensure reliability in agent-mediated clinical pipelines.


\bibliographystyle{splncs04}
\bibliography{main}

\end{document}